\documentclass[a4paper,11pt]{article}
\usepackage{pos}

\title{CP violation in non-leptonic $B$ decays as a portal to New Physics}
\author*[a,b]{Robert Fleischer}
\affiliation[a]{Nikhef, Science Park 105, NL-1098 XG Amsterdam, Netherlands
}

\affiliation[b]{Department of Physics and Astronomy, Vrije Universiteit Amsterdam
NL-1081 HV Amsterdam, Netherlands}

\emailAdd{robert.fleischer@nikhef.nl}
\abstract{CP violation offers a powerful probe for testing the Standard Model. In this endeavour, non-leptonic decays of B mesons play a key role. I will discuss benchmark decays and puzzling patterns emerging from theoretical analyses of the current data.}

\FullConference{%
  7th Symposium on Prospects in the Physics of Discrete Symmetries (DISCRETE 2020-2021)\\
  29th November - 3rd December 2021\\
 Bergen, Norway}

\begin{document}
\maketitle

\section{Setting the stage}
In the Standard Model (SM), the Cabibbo--Kobayashi--Maskawa (CKM) matrix gives rise to a plethora of quark-flavour physics phenomena and allows us to accommodate CP violation through a complex phase. The ``quark-flavour code" of the SM is encoded in weak decays of $K$, $D$ and $B$ mesons, where the latter offer particularly interesting probes. In scenarios for New Physics (NP), we typically have new sources for flavour and CP violation. Despite its tremendous success, we have indications that the SM cannot be complete, where key examples are the baryon asymmetry of the Universe, as well as the established non-vanishing neutrino masses and dark matter. Furthermore, we do not understand the origin of the structure of the SM and the patterns of masses and flavour-mixing parameters with their intriguing hierarchies at a more fundamental level.

Where do we stand? Data taken at the Large Hadron Collider (LHC) have led to the discovery of the Higgs boson by the ATLAS and CMS collaborations in 2012. These experiments are further impressively studying the properties of this fascinating particle. However, no other new particles and deviations from the SM were so far seen by these experiments at the high-energy frontier. Concerning CP violation and the quark-flavour sector, we have a globally consistent picture with the CKM paradigm. However, at the high-precision frontier, various ``anomalies" with respect to the SM picture have emerged in observables of certain $B$-meson decays, where semi-leptonic channels are in the spotlight since a few years, with potential signals of the violation of lepton flavour universality, as discussed in detail in \cite{Altmannshofer}. However, also non-leptonic $B$ decays show puzzling effects, which will be the focus of this presentation. Further long-standing tensions with respect to the SM arise in the anomalous magnetic moment of the muon. 

In view of this situation, we may conclude that NP described by a Lagrangian
\begin{equation}
{\cal L}={\cal L}_{\rm SM} + {\cal L}_{\rm NP}(\varphi_{\rm NP}, g_{\rm NP}, m_{\rm NP}, ...)
\end{equation}
is characterised by a large NP scale $\Lambda_{\rm NP}$ far beyond the TeV regime, 
which would be challenging for direct searches at ATLAS and CMS, or/and symmetries prevent large NP effects in flavour-changing neutral currents and the flavour sector. Fortunately, we are facing exciting future prospects, in particular due to the LHC upgrade(s) and the data taking at the Belle II experiment. 

In order to deal with NP effects entering far beyond the electroweak scale, ``Effective Field Theories" offer a powerful framework. Here 
 the heavy degrees of freedom, i.e.\ NP particles as well as the top quark and the $Z$ and $W$ bosons of the SM, are 
``integrated out" from appearing explicitly for low-energy phenomena, 
and are then described by short-distance functions. In recent decades, perturbative 
QCD corrections were calculated and renormalisation group techniques applied for the summation of large logarithms. Such analyses have been performed for the SM, but also a wide spectrum of specific NP scenarios. In order to perform model-independent analyses of NP effects, Standard Model Effective Theories (SMEFT) offer an interesting tool.

Utilising these techniques, for the description of non-leptonic $B$-meson decays $\overline{B}\to \overline{f}$, 
low-energy effective Hamiltonians are used that take the following structure in the SM:
\begin{equation}
\langle \overline{f}|{\cal H}_{\mbox{{\scriptsize eff}}}|\overline{B}\rangle=
\frac{G_{\rm F}}{\sqrt{2}}\sum_{j}\lambda_{\rm CKM}^{j}\sum_{k}
C_{k}(\mu) \langle \overline{f}|Q^{j}_{k}(\mu)|\overline{B}\rangle.
\end{equation}
Here $G_{\rm F}$ is Fermi's constant, the $\lambda_{\rm CKM}^j$ denote products of CKM matrix elements, $\mu$ is a renormalization scale, while the $Q^{j}_{k}$ are four-quark operators with their short-distance Wilson coefficients $C_{k}(\mu)$. Such Hamiltonians characterise different quark-flavour processes, and specific decays are described through the corresponding hadronic matrix elements. Whereas the Wilson coefficients can be calculated, including QCD corrections leading to the $\mu$ dependences, the non-perturbative
hadronic matrix elements usually are affected by large hadronic uncertainties. Beyond the SM, the Wilson coefficients of SM operators may get new contributions but also new operators may arise. In general, we may also get new sources of CP violation, which are
described by complex couplings. 

Non-leptonic $B$ decays are key players in the field of CP violation. The reason is that CP-violating rate asymmetries are generated through interference effects which may arise in such processes. Theoretical predictions are affected by hadronic matrix elements of four-quark operators. Fortunately, it is possible to circumvent the calculation of the hadronic matrix elements $\langle\overline{f}|Q_k^j(\mu)|\overline{B}\rangle$ in studies of CP violation: First, amplitude relations can be exploited, either in exact relations, involving pure ``tree" decays of the kind $B\to D K$, or approximate relations, which follow from flavour symmetries of strong interactions, i.e.\ $SU(2)$ isospin or flavour $SU(3)$, involving $B\to\pi\pi$, $B\to\pi K$ and $B_{(s)}\to KK$ modes. Second, in decays of 
neutral $B_d$ or
$B_s$ mesons, we may get interference effects through $B^0_q$--$\overline{B}^0_q$ mixing ($q=d,s$) should both the $B^0_q$ and 
the $\overline{B}^0_q$ mesons decay into the same final state, thereby leading to ``mixing-induced" CP violation. If one CKM amplitude 
dominates the decay, the corresponding hadronic matrix elements cancel in such ``mixing-induced" CP asymmetries, while ``direct" CP violation -- arising directly at the decay amplitude level through interference between different decay contributions -- would vanish.

Measurements of CP violation in non-leptonic $B$ decays allow determinations of the angles of the Unitarity Triangle (UT) of the CKM matrix. For detailed analyses of the UT, see Refs.~\cite{UT-1}. In the presence of NP contributions, discrepancies should emerge between constraints from various processes. A key ingredient for such studies is given by the $B^0_q$--$\overline{B}^0_q$ mixing phases.

\boldmath
\section{Precision measurements of the $B^0_q$--$\overline{B}^0_q$ mixing phases}
\unboldmath
In the SM, $B^0_q$--$\overline{B}^0_q$ mixing arises from box topologies \cite{Nierste}. This phenomenon is very sensitive to possible NP contributions which may either enter the loop processes or at the tree level, as, for instance, in models with extra $Z'$ bosons. In general, such effects involve also new sources for CP violation. The CP-violating phases associated with $B^0_q$--$\overline{B}^0_q$ mixing can be written as follows:
\begin{equation}\label{phi-q}
\phi_d=2\beta+\phi_d^{\rm NP}, \quad \phi_s=-2\lambda^2\eta +\phi_s^{\rm NP},
\end{equation}
where $\beta$ is one of the angles of the UT, $\lambda=|V_{us}|\approx 0.22$ and $\eta$ are CKM parameters, and the 
$\phi_q^{\rm NP}$ denote NP phases. In our quest for physics beyond the SM, we are moving towards new frontiers. It is crucial for resolving smallish NP effects to have a critical look at theoretical analyses and their approximations, where strong interactions lead to hadronic uncertainties. The goal is to match the experimental and theoretical precisions. Benchmark decays for exploring CP violation are $B^0_d\to J/\psi K_{\rm S}$ and $B_s^0\to J/\psi \phi$, which allow measurements of the $B^0_{d,s}$--$\bar B^0_{d,s}$ mixing phases $\phi_{d,s}$. Uncertainties arise from doubly Cabibbo-suppressed penguin contributions. However, these effects are usually neglected. How big are these contributions and how can they be controlled?

In the SM, the  amplitude of the decay $B_d^0\to J/\psi\, K_{\rm S}$ takes the following form \cite{RF-99}:
\begin{equation}\label{ABdpsiKS}
A(B_d^0\to J/\psi\, K_{\rm S})=\left(1-\lambda^2/2\right){\cal A'}
\left[1+ \epsilon a'e^{i\theta'}e^{i\gamma}\right], 
\end{equation}
where ${\cal A'}$ and $a'e^{i\theta'}$ are hadronic parameters which are governed by colour-suppressed tree and penguin topologies, respectively, $\gamma$ is the corresponding UT angle, and $\epsilon\equiv\lambda^2/(1-\lambda^2)\sim0.05$. The mixing-induced and direct CP asymmetries $S$ and $C$, respectively, satisfy the following relation:
\begin{equation}\label{CP-rel}
\frac{S(B_d\to J/\psi K_{\rm S})}{\sqrt{1-C(B_d\to J/\psi K_{\rm S})^2}}
=\sin(\phi_d+\Delta\phi_d),
\end{equation}
where $\Delta\phi_d$ is a hadronic phase shift \cite{FFJM}:
\begin{equation}
\sin\Delta\phi_d \propto 2 \epsilon a'\cos\theta' \sin\gamma+\epsilon^2a'^2, \quad
\cos\Delta\phi_d \propto 1+ 2 \epsilon a'\cos\theta' \cos\gamma+\epsilon^2a'^2
\cos2\gamma.
\end{equation}
Neglecting the doubly Cabibbo-suppressed $\epsilon a'$ penguin parameters, we obtain the well-known result $S(B_d\to J/\psi K_{\rm S})=\sin\phi_d$, which is usually applied to determine the UT angle $\beta$ from the measured CP-violating asymmetries. 

The decay $B^0_s\to J/\psi\phi$ arises from the same quark-level processes as the $B_d^0\to J/\psi\, K_{\rm S}$ channel. However, 
the final state with two vector mesons is a mixture of CP-odd and CP-even eigenstates $f=0,\parallel$ and $\perp$, respectively.
In order to disentangle them, an angular analysis of the $J/\psi \to\mu^+\mu^-$, $\phi \to\ K^+K^-$ decay products 
has to be performed in the time-dependent decay rate analysis \cite{DDF,DFN,dBF}. In analogy to $B_d^0\to J/\psi\, K_{\rm S}$, doubly Cabibbo-suppressed penguins effects lead to the following effective CP-violating mixing phase entering the CP-violating 
observables \cite{FFM}:
\begin{equation}
\phi_{s,(\psi \phi)_f}^{\rm eff} = \phi_s+\Delta\phi_s^{(\psi \phi)_f} \equiv  \phi_s+\Delta\phi_s^{f}.
\end{equation}
For a smallish mixing phase $\phi_s$ in the few degree regime, as follows from the picture of the experimental data, even a 
hadronic phase shift $\Delta\phi_s^{f}$ at the $1^\circ$ level would have a significant impact.

\begin{figure}[t] %  figure placement: here, top, bottom, or page
   \centering
   \includegraphics[width=6.5truecm]{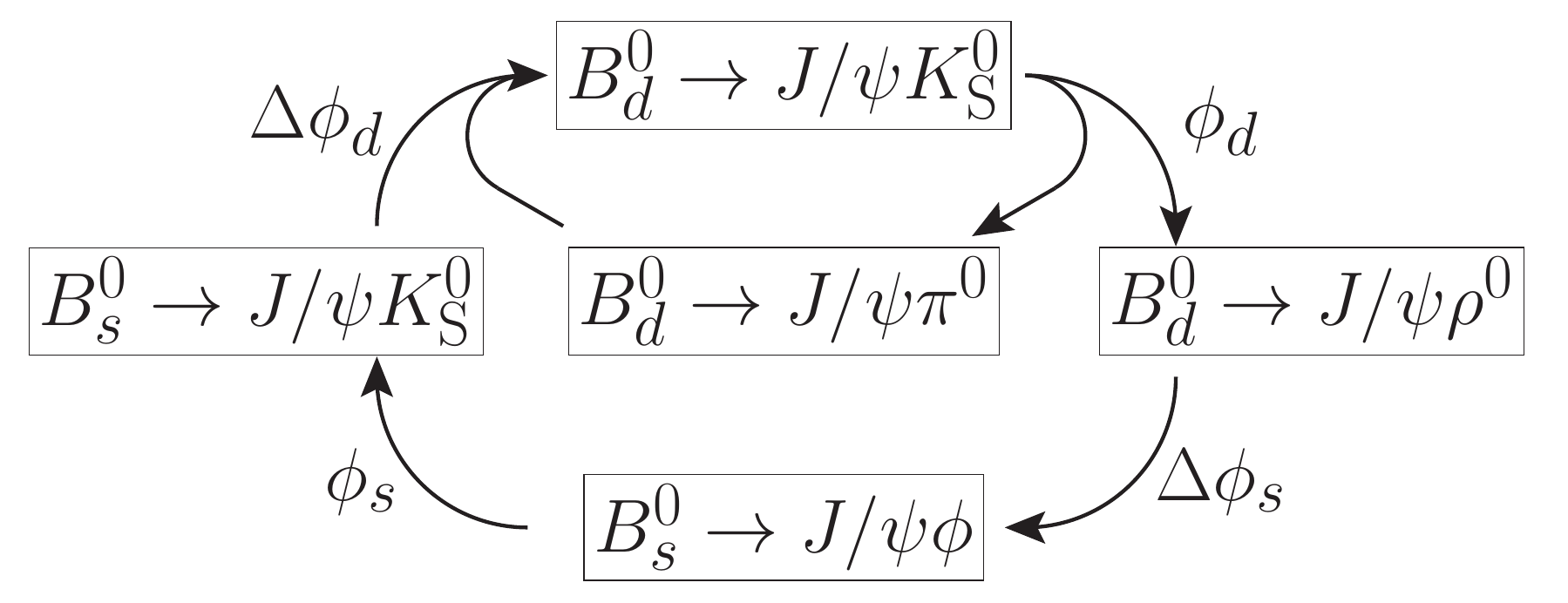} 
   \caption{Illustration of the interplay between the determination of the CP-violating phases $\phi_d$ and $\phi_s$ and their hadronic penguin shifts $\Delta\phi_d$ and $\Delta\phi_s$ through control channels (from Ref.~\cite{BdBFM}).}
   \label{fig:interplay}
\end{figure}

In the era of Belle II and the LHCb upgrades, the experimental precision requires the control of the penguin corrections to 
reveal possible CP-violating NP contributions to $B^0_q$--$\bar B^0_q$ mixing. The topic receives long-standing interest in the 
theory community (see, e.g., Refs.~\cite{RF-99,FFJM,FFM,CPS,GR-08,jung,FNW,dBF,BdBFM}). The hadronic phase shifts 
$\Delta\phi_d$ and $\Delta \phi_s^f$ cannot be reliably calculated within QCD. However, we may 
use ``control channels" $B^0_s\to J/\psi K_{\rm S}$, $B^0_d\to J/\psi \pi^0$, and $B^0_s\to J/\psi \rho^0$, which have a different CKM
amplitude structure with the key feature that the penguin parameters are not suppressed by $\epsilon$. This can nicely be seen 
in the case of the $B^0_s\to J/\psi K_{\rm S}$ amplitude \cite{RF-99}:
\begin{equation}
A(B^0_s\to J/\psi K_{\rm S})\propto\left[1-a e^{i\theta}e^{i\gamma}\right],
\end{equation}
which should be compared with Eq.~(\ref{ABdpsiKS}). Measurements of the CP-violating $B^0_s\to J/\psi K_{\rm S}$ asymmetries allow
the determination of $a$ and $\theta$, which can then be related to their $B^0_d\to J/\psi K_{\rm S}$ counterparts through the 
$U$-spin symmetry of strong interactions, $a e^{i\theta}=a' e^{i\theta'}$, thereby allowing us to take these effects into account in the determination of $\phi_d$ from Eq.~(\ref{CP-rel}).

In Ref.~\cite{BdBFM}, a simultaneous strategy of various control channels was proposed and applied to the currently available data, utilising the $SU(3)$ flavour symmetry of strong interactions. The point is that there is a subtle interplay between the mixing phases and decays, as illustrated in Fig.~\ref{fig:interplay}. For a detailed discussion, also of the numerical analysis with correlation plots, the reader is referred to that paper and the recent 
update in Ref.~\cite{BdBFM-CKM}. Let us here just give the main numerical results:
\begin{equation}
a=0.14^{+0.17}_{-0.11}, \quad \theta=\left(173^{+35}_{-45} \right)^\circ,  \quad
\phi_d=\left(44.4^{+1.6}_{-1.5} \right)^\circ,
\end{equation}
which should be compared with the measured value  $\phi_{d,J/\psi K^0}^{\rm eff}=\left(43.6\pm1.4\right)^\circ$, and
\begin{equation}\label{BpsiV}
a_V=0.044^{+0.0.085}_{-0.038}, \quad \theta_V=\left(306^{+48}_{-112} \right)^\circ,  \quad
\phi_s=-\left(4.2\pm1.4 \right)^\circ,
\end{equation}
which should be compared with $\phi_{s,J/\psi\phi}^{\rm eff}=-\left(4.1\pm1.3\right)^\circ$. In the analysis of the 
$B_{(s)}\to J/\psi V$ modes, polarisation-dependent effects had to be ignored due to the current lack of data. In the future, it would be important to make polarisation-dependent measurements, which could then be implemented in the strategy to further refine the analysis. The results in Eq.\ (\ref{BpsiV})
correspond to the penguin shift $\Delta\phi_s= \left(0.14^{+0.54}_{-0.70}\right)^\circ$. Using Eq.~(\ref{phi-q}) with 
$\phi_s^{\rm SM}=-2\lambda^2\eta=-\left(2.01\pm0.12 \right)^\circ$ yields $\phi_s^{\rm NP}=-\left(2.2\pm1.4\right)^\circ$. The future scenarios studied in Ref.~\cite{BdBFM-CKM} show that CP-violating NP effects could be revealed in the high-precision era 
with $5\sigma$ significance. On the other hand, for the $B_d$-meson system, the SM prediction of $\phi_d$ is a limiting factor, thereby
making $\phi_d$ less favourable. 

The strategy presented in Ref.~\cite{BdBFM} provides also interesting insights into factorisation. Having the hadronic penguin parameters at hand and using information from semileptonic $B$ decays to minimise the impact of hadronic form factors, 
effective colour-suppression factors $a_2$ can be determined from the data, showing agreement with the picture of naive 
factorisation. This is an interesting finding as factorisation is -- a priori -- not expected to work well in these colour-suppressed decays. Using these results, non-factorisable $SU(3)$-breaking effects can be constrained at the $5\%$ level, thereby showing that the
method to control the penguin effects through control channels is robust.

\boldmath
\section{The $B^0_s\to D_s^\mp K^\pm$ puzzle}
\unboldmath
In the SM, the decays $\overline{B}^0_s\to D_s^+K^-$ and $B^0_s\to D_s^+K^-$ arise from colour-allowed tree topologies. Since both 
$\overline{B}^0_s$ and $B^0_s$ mesons may decay into the same final state $D_s^+K^-$, we obtain CP violation through interference 
effects between the decay paths that are due to the $B^0_s$--$\overline{B}^0_s$ oscillations \cite{ADK,RF-BsDsK,dBFKMST}. 
The corresponding observables arise in the time-dependent rate asymmetry
\begin{equation}
\frac{\Gamma(B^0_s(t)\to D_s^{+} K^-) - \Gamma(\bar{B}^0_s(t)\to D_s^{+} K^-) }
	{\Gamma(B^0_s(t)\to D_s^{+} K^-) + \Gamma(\bar{B}^0_s(t)\to D_s^{+} K^-) }  
	= \frac{{C}\,\cos(\Delta M_s\,t) + {S}\,\sin(\Delta M_s\,t)}
	{\cosh(y_s\,t/\tau_{B_s}) + {\cal A}_{\Delta\Gamma}\,\sinh(y_s\,t/\tau_{B_s})},
\end{equation}
where $\Delta M_s$ is the mass difference of the $B_s$ mass eigenstates and 
$y_s\equiv \Delta\Gamma_s/(2\,\Gamma_s)=0.062 \pm 0.004$ characterises their decay width difference. A similar expression, with 
observables $\overline{C}$, $ \overline{S}$ and $\overline{{\cal A}}_{\Delta\Gamma}$, 
holds for the decays into the CP-conjugate final state $D_s^-K^+$. The quantities
\begin{equation}
 C=\frac{1-|\xi|^2}{1+|\xi|^2},  \quad S= \frac{2\,{\rm Im}{\,\xi}}{1 + |\xi|^2}, \quad 
 \mathcal{A}_{\Delta \Gamma}=\frac{2\,{\rm Re}\,\xi}{1+|\xi|^2}
\end{equation}
and their CP-conjugates can be extracted from the time-dependent rate asymmetries, allowing the determination of the observables $\xi$ and $\overline{\xi}$. In their product, the hadronic parameters cancel \cite{RF-BsDsK}:
\begin{equation}\label{xi-prod-SM}
{\xi} \times \bar{\xi}= e^{-i2( \phi_s + \gamma)} .
\end{equation}
Consequently, the CP-violating phase $\phi_s+\gamma$ can be determined in a theoretically clean way \cite{ADK,RF-BsDsK}. 
Since $\phi_s$ is determined through $B^0_s\to J/\psi \phi$ and similar modes, as discussed in the previous section, the UT 
angle $\gamma$ can be extracted.  

In Ref.~\cite{LHCb-BsDsK}, the LHCb collaboration reported an experimental analysis of CP violation in
the $B^0_s\to D_s^\mp K^\pm$ system, 
finding the result $\gamma=\left(128^{+17}_{-22}\right)^\circ$ (mod $180^\circ$). Here the SM relation $C+\overline{C}=0$ was assumed. The result for $\gamma$ is puzzling since analyses of the UT and other $\gamma$ determinations using pure tree decays give values in the $70^\circ$ regime \cite{UT-1,LHCb:2021dcr}. 

This intriguing situation has recently been studied in detail in Refs.~\cite{FM-1,FM-2}. Using
\begin{equation}
\tan(\phi_s+\gamma)=-\left[\frac{\overline{S}+ S}{ \mathcal{\overline{A}}_{\Delta \Gamma}+ \mathcal{A}_{\Delta \Gamma}}\right], 
\quad
\tan\delta_s= \left[\frac{\overline{S} - S}{ \mathcal{\overline{A}}_{\Delta \Gamma}+ \mathcal{A}_{\Delta \Gamma}}\right],
\end{equation}
where $\delta_s$ is the CP-conserving strong phase difference between the $\overline{B}^0_s\to D_s^+K^-$ and $B^0_s\to D_s^+K^-$ decay amplitudes, a transparent determination of these parameters is possible. The corresponding analysis gives a picture in full agreement with the complex LHCb fit. The solutions modulo $180^\circ$ can actually be excluded, since the corresponding $\delta_s$ around $180^\circ$ would be in conflict with factorisation, while $\delta_s= (-2^{+13}_{-14})^\circ$ is in excellent agreement with this
theoretical framework.

How could NP effects enter this measurement? They could give rise to new CP-violating contributions to $B^0_s$--$\overline{B}^0_s$ mixing, thereby affecting $\phi_s$. However, such effects are included as this phase is determined through 
$B^0_s\to J/\psi \phi$ and penguin control modes. Using the corresponding value taking penguin corrections into account shifts 
the LHCb result to $\gamma=\left(131^{+17}_{-22}\right)^\circ$. 

Consequently, this puzzling value of $\gamma$ would require NP contributions -- with new sources of CP violation -- at the 
decay amplitude level of the $B^0_s\to D_s^\mp K^\pm$ system. Such effects should manifest themselves also 
in the branching ratios of the corresponding decays. Concerning the branching ratios of $B_s$ decays, there are subtleties due to 
$B^0_s$--$\overline{B}^0_s$ mixing \cite{dBFKMST,BR-paper}. The ``theoretical" branching ratios refer to a situation where the
mixing effects are ``switched off":
\begin{equation}
\mathcal{B}_{\rm{th}} \equiv \frac{1}{2}\left[ \mathcal{B}(\bar{B}^0_s \rightarrow D_s^+ K^-)_{\rm{th}} +  
\mathcal{B}({B}^0_s \rightarrow D_s^+ K^-)_{\rm{th}}\right].
\end{equation}
The observable $\xi$ allows us to disentangle the decay paths (in analogy for $D_s^-K^+$):
\begin{equation}
\mathcal{B}(\bar B^0_s\to D_s^+K^-)_{\rm{th}}=2 \left(\frac{|\xi|^2}{1+|\xi|^2} \right)\mathcal{B}_{\rm{th}}, \quad
\mathcal{B}(B^0_s\to D_s^+K^-)_{\rm{th}}=2 \left(\frac{1}{1+|\xi|^2} \right)\mathcal{B}_{\rm{th}}.
\end{equation}
The ``experimental" branching ratios refer to the following time-integrated rates:
\begin{equation}
\mathcal{B}_{\rm {exp}} = 
 \frac{1}{2} \int_0^{\infty} \! \left[\Gamma (\bar{B}_s^0 (t)\rightarrow D_s^+ K^-) + \Gamma (B_s^0 (t)\rightarrow D_s^+ K^-) \right]
 \mathrm{d}t ,
\end{equation}
and are related to the theoretical branching ratios through
\begin{equation}
\mathcal{B}_{\rm{th}} = \left[ \frac{1-y_s^2}{1+ \mathcal{A}_{\Delta \Gamma_s} y_s} \right] \mathcal{B}_{\rm{exp}}.
\end{equation}
Unfortunately, only a measurement of the following average is available:
\begin{equation}
\mathcal{B}^{\rm{exp}}_\Sigma \equiv \mathcal{B}_{\rm{exp}} + \bar{\mathcal{B}}_{\rm{exp}} \equiv 2\, \langle\mathcal{B}_{\rm{exp}}\rangle = (2.27 \pm 0.19) \times 10^{-4}.
\end{equation}
Assuming the SM, as the LHCb collaboration, yields
\begin{equation}
 \mathcal{B}_{\rm{th}} =  \bar{\mathcal{B}}_{\rm{th}}=
\left[\frac{1-y_s^2}{1+y_s\langle {\cal A}_{\Delta\Gamma}\rangle_+}\right]\langle\mathcal{B}_{\rm{exp}}\rangle \quad\mbox{with}\quad
\langle \mathcal{A}_{\Delta \Gamma} \rangle_+\equiv
\frac{\mathcal{\overline{A}}_{\Delta \Gamma}+ \mathcal{A}_{\Delta \Gamma}}{2}.
\end{equation}
Finally, the following branching ratios can be extracted from the data:
\begin{equation}
\mathcal{B}(\bar{B}^0_s \rightarrow D_s^{+}K^{-})_{\rm th}=(1.94 \pm 0.21) \times 10^{-4}, \quad
 \mathcal{B}(B^0_s \rightarrow D_s^{+}K^{-})_{\rm th}=(0.26 \pm 0.12) \times 10^{-4}.
\end{equation}

\begin{figure}[t] %  figure placement: here, top, bottom, or page
   \centering
   \includegraphics[width=5.5truecm]{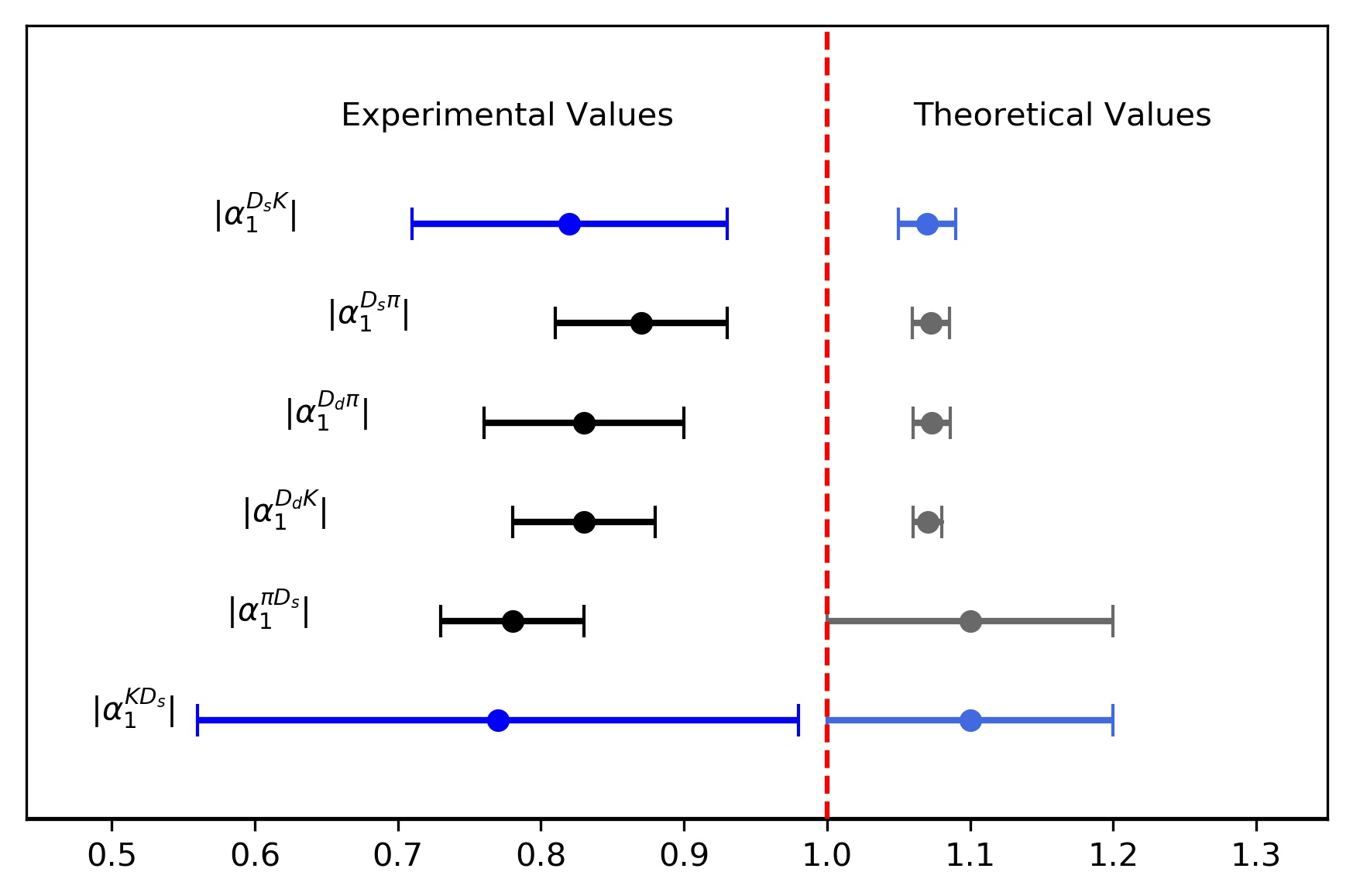} 
   \caption{Experimental and SM values of the $|a_1|$ parameters for various decay processes \cite{FM-1,FM-2}.}\label{fig:a1}
\end{figure}

The framework for the interpretation of these quantities is provided by factorisation, which is well supported 
through the measured $\delta_s$. In order to minimise the impact of
hadronic form factor uncertainties, it is useful to introduce ratios with respect to semileptonic $B$ decays \cite{FM-1,FM-2}:
\begin{equation}
  R_{D_s^{+}K^{-}}\equiv\frac{\mathcal{B}(\overline{B}^0_s \rightarrow D_s^{+}K^{-})_{\rm th}}{{\mathrm{d}\mathcal{B}\left(\overline{B}^0_s \rightarrow D_s^{+}\ell^{-} \bar{\nu}_{\ell} \right)/{\mathrm{d}q^2}}|_{q^2=m_{K}^2}} = 
  6 \pi^2 f_{K}^2 |V_{us}|^2   X_{D_s K} |a_{\rm 1 \, eff }^{D_s K}|^2,
\end{equation}
where $ f_{K}$ is the kaon decay constant, $V_{us}$ denotes the corresponding CKM matrix element, and $X_{D_s K}$ is a calculable quantity governed by phase-space effects. In the parameter
\begin{equation}
a_{\rm 1 \, eff }^{D_s K}=a_{1}^{D_s K} \left(1+\frac{E_{D_s K}}{T_{D_s K}}\right),
\end{equation}
where 
$a_{1}^{D_s K}$ characterizes factorisation of the colour-allowed tree amplitude $T_{D_s K}$, which is a ``show-case" example for factorization predicting  $|a_1^{D_sK}| = 1.07\pm0.02$ \cite{Beneke:2000ry,Huber:2016xod,Beneke:2021jhp}, 
while the non-factorisable exchange amplitude $E_{D_s K}$ is constrained 
through data as $|1+E_{D_s K}/T_{D_s K}|=1.00\pm0.08$. Finally, the experimental data give $|a_{\rm 1}^{D_s K}| = 0.82 \pm 0.11$,
which is significantly smaller than the QCD factorisation prediction. Interestingly, the data give a similar pattern for the 
$\overline{B}^0_s \rightarrow K^+ D_s^-$ channel, with $|a_{\rm 1}^{K D_s}| =0.77 \pm 0.19$. Consequently, this puzzling finding complements the intriguing pattern for $\gamma$ arising from the CP-violating observables, as would be suggested by CP-violating NP contributions entering the corresponding decay amplitudes. This result is even more exciting since analogous patterns arise in decays with similar dynamics as summarised in Fig.~\ref{fig:a1}, where $\bar B^0_d\to D_d^+ K^-$ stands out with a discrepancy
of $4.8 \,\sigma$. Puzzlingly small branching ratios for this mode and the $\bar B^0_d\to D_d^+ \pi^-$, $\bar B^0_s\to D_s^+ \pi^-$ 
decays were also noted in Refs.\ \cite{FST-BR,Bordone:2020gao}, and have been analysed within NP physics beyond the SM 
\cite{Iguro:2020ndk,Cai:2021mlt,Bordone:2021cca}. The interesting possibility of NP effects in non-leptonic tree-level decays of $B$ mesons was discussed in detail in Refs.\ \cite{Brod:2014bfa,Lenz:2019lvd}. 

\begin{figure}[t] %  figure placement: here, top, bottom, or page
   \centering
   \includegraphics[width=4.5truecm]{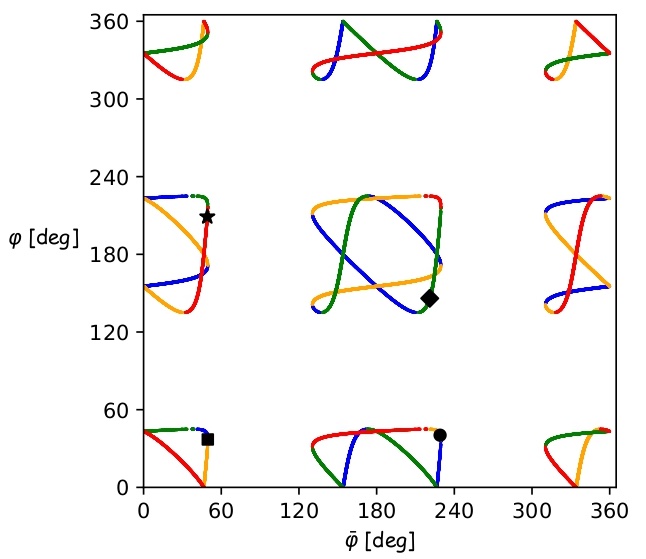} 
    \includegraphics[width=4.0truecm]{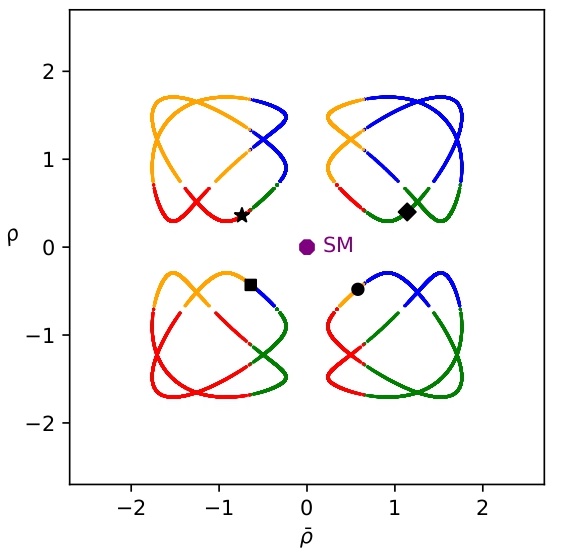} 
   \caption{Correlations between the NP parameters of the $B^0_s\to D_s^\mp K^\pm$  system \cite{FM-1,FM-2}.}\label{fig:NP-corr}
\end{figure}

In order to take NP effects into account, the $\overline{B}^0_s \rightarrow D_s^+ K^-$ amplitude can be generalised as 
\begin{equation}
A(\overline{B}^0_s \rightarrow D_s^+ K^-) = A(\overline{B}^0_s \rightarrow D_s^+ K^-)_{{\rm{SM}}} \left[ 1 + \bar{\rho} \, e^{i \bar{\delta}}
e^{+i \bar{\varphi}} \right], 
\quad
\bar{\rho} \, e^{i \bar{\delta}} e^{i \bar{\varphi}}  \equiv \frac{ A(\bar{B}^0_s \rightarrow D_s^+ K^-)_{{\rm{NP}}} }{  A(\bar{B}^0_s \rightarrow D_s^+ K^-)_{{\rm{SM}}} },
\end{equation}
where $\bar{\delta}$ and $\bar{\varphi}$ are CP-conserving and CP-violating phases, respectively. An analogous expression holds for 
the $\overline{B}^0_s \rightarrow D_s^- K^+$ channel. The generalisation of the SM relation (\ref{xi-prod-SM}) is given as follows:
\begin{equation}
\xi \times \bar{\xi}  = \sqrt{1-2\left[\frac{C+\bar{C}}{\left(1+C\right)\left(1+\bar{C}\right)}
\right]}e^{-i\left[2 (\phi_s +\gamma_{\rm eff})\right]},
\end{equation}
with $\gamma$ entering through the ``effective" angle 
\begin{equation}
\gamma_{\rm eff}\equiv \gamma+\frac{1}{2}\left(\Delta\Phi+\Delta\bar{\Phi}\right)=
\gamma-\frac{1}{2}\left(\Delta\varphi+\Delta\bar{\varphi}\right),
\end{equation}
where the NP phase shifts can be expressed in terms of $\bar{\rho}$, $\bar{\delta}$, $\bar{\varphi}$ and their CP conjugates as discussed in detail in Refs.~\cite{FM-1,FM-2}. Using the experimental information encoded in the branching ratios and CP asymmetries, the constraints on the NP parameters shown in Fig.~\ref{fig:NP-corr} can be obtained for the central values of the observables. 
Taking also the uncertainties into account, NP amplitudes in the (30-50)\% range of the SM amplitudes could accommodate the 
current experimental data.

\boldmath
\section{The $B\to\pi K$ puzzle}
\unboldmath
Interestingly, $B\to \pi K$ decays are dominated by QCD penguin topologies since $\overline{b}\to \overline{u}u\overline{s}$ tree topologies are strongly CKM suppressed with respect to the penguin contributions. Electroweak (EW) penguins allow us to divide the $B\to\pi K$ system into two classes: In the $B^0_d\to \pi^-K^+$ and $B^+\to\pi^+K^0$ decays, EW penguins are colour-suppressed and hence play a minor role. On the other hand, in the $B^0_d\to \pi^0K^0$ and $B^+\to\pi^0K^+$ modes, EW penguins may enter in 
colour-allowed form and therefore play a sizeable role, even competing with tree contributions \cite{RF-94,GHLR-EWP,RF-95,BFRS}.

A particularly interesting situation arises in the $B^0_d\to \pi^0K_{\rm S}$ decay \cite{FJPZ,FJV,FJMV}, which offers the following 
CP-violating time-dependent rate asymmetry:
\begin{equation}
\frac{\Gamma(\overline{B}_d^0(t) \rightarrow \pi^0K_{\rm S}) - 
\Gamma(B_d^0(t) \rightarrow \pi^0K_{\rm S})}{\Gamma(\overline{B}_d^0(t) \rightarrow \pi^0K_{\rm S}) + 
\Gamma(B_d^0(t) \rightarrow \pi^0K_{\rm S})}=
A^{\pi^0K_{\rm S}}_{\rm CP} \cos(\Delta M_dt) + S^{\pi^0K_{\rm S}}_{\rm CP}\sin(\Delta M_dt),
\end{equation}
where $A^{\pi^0K_{\rm S}}_{\rm CP} $ and $S^{\pi^0K_{\rm S}}_{\rm CP}$ are the direct and mixing-induced CP 
asymmetries, respectively. 
The isospin symmetry of strong interactions implies the relation
\begin{equation}\label{iso-rel}
\sqrt{2} A(B^0_d \to \pi^0 K^0) + A(B^0_d \to \pi^- K^+)   \equiv 3A_{3/2},
\end{equation}
where the $I=3/2$ isospin amplitude $A_{3/2}$ takes the form
\begin{equation}\label{iso-rel-2}
3A_{3/2} =  - (\hat{T} +\hat{C})\left(e^{i\gamma} - q e^{i\phi} e^{i\omega}\right).
\end{equation}
Here the absolute value of the sum of the colour-allowed and colour-suppressed tree amplitudes $\hat{T}$ and $\hat{C}$, respectively, can be determined with the help of the $B^+\to\pi^+ \pi^0$ branching ratio:
\begin{equation}\label{RTC}
	|\hat{T}+\hat{C}| = R_{T+C}\left|V_{us}/V_{ud}\right|\sqrt{2} |A(B^+\to\pi^+ \pi^0)|,
\end{equation}
where $R_{T+C}\sim f_K/f_\pi$ describes $SU(3)$-breaking corrections. The EW penguins are described by
\begin{equation} \label{q}
	q e^{i\phi} e^{i\omega} \equiv - \left[\frac{\hat{P}_{EW} + \hat{P}_{EW}^{\rm C}}{\hat{T} +\hat{C}}\right] =
	\frac{-3}{2\lambda^2 R_b}\left[\frac{C_9 + C_{10}}{C_1 + C_2}\right] R_q = (0.68 \pm 0.05) R_q,
\end{equation}
where $\hat{P}_{EW}$ and $\hat{P}_{EW}^{\rm C}$ are colour-allowed and colour-suppressed amplitudes, respectively, $R_b$ is the side of the UT from the origin, and $R_q$ describes $SU(3)$-breaking corrections which can be calculated using QCD factorisation techniques and input 
from lattice QCD \cite{FJPZ}. The $C_{1,2}$ and $C_{9,10}$ are short-distance Wilson coefficient functions of current--current and EW
penguin operators, respectively. Applying the SM values of these coefficients gives the numerical value in Eq.~(\ref{q}).

\begin{figure}[t] %  figure placement: here, top, bottom, or page
   \centering
   \includegraphics[width=4.5truecm]{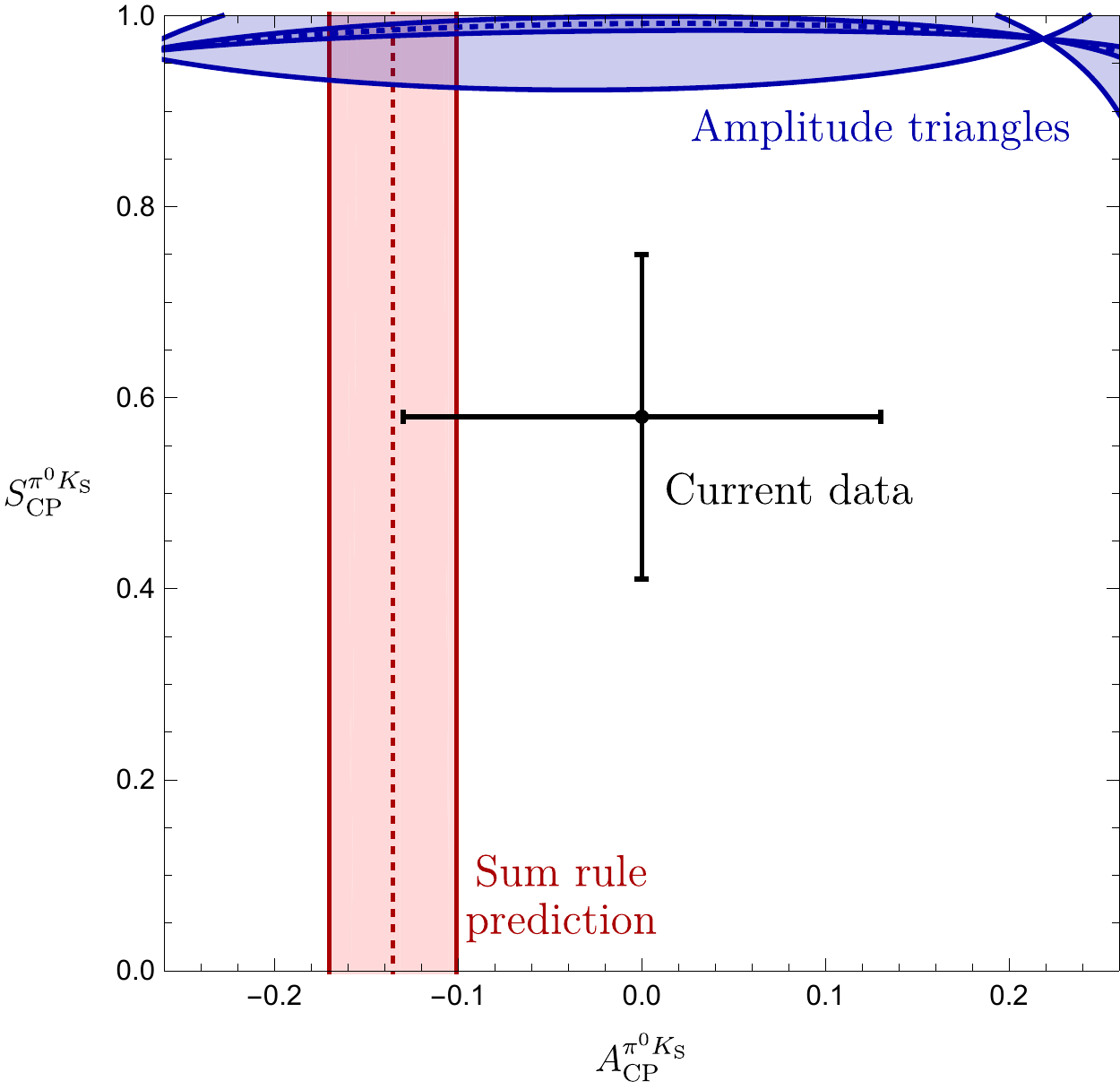} 
   \caption{Correlation between the CP asymmetries of $B^0_d\to \pi^0K_{\rm S}$, as discussed in the text (from \cite{FJMV}).}\label{fig:A-S-corr}
\end{figure}

Introducing the angle $\phi_{00} \equiv \rm{arg}(\bar{A}_{00} A^*_{00})$ between the decay amplitude $A_{00} \equiv A(B_d^0 \to \pi^0 K^0)$ and its CP-conjugate $\bar{A}_{00}$, the following relation can be derived: 
\begin{equation} 
	S^{\pi^0K_{\rm S}}_{\rm CP} = \sin(\phi_d - \phi_{00})\sqrt{1- (A^{\pi^0K_{\rm S}}_{\rm CP})^2}.
\end{equation}
Using the isospin relation in Eq.~(\ref{iso-rel}) with the minimal $SU(3)$ input entering Eqs.~(\ref{RTC}) and (\ref{q}), this expression allows the calculation of correlations between the direct and mixing-induced CP asymmetries of the $B^0_d\to \pi^0K_{\rm S}$ decay.
Using the current data results in the ``Isospin relation" upper band in Fig.~\ref{fig:A-S-corr}. The vertical ``Sum rule prediction" band is related to sum rules involving branching ratios and direct CP asymmetries of $B\to\pi K$ decays \cite{gro,Gro-Ro}. Consequently, we encounter yet another puzzling pattern related to CP violation in non-leptonic $B$ decays  \cite{FJPZ,FJV,FJMV}. 
Using further information from charged $B\to\pi K$ decays, which satisfy also an isospin relation in analogy to Eq.~(\ref{iso-rel}),
correlations in the $\phi$--$q$ plane of the EW penguin parameters can be obtained, as discussed in detail in Ref.~\cite{FJMV}. NP effects with new sources for CP violation entering through such topologies could accommodate the current data.

\section{Conclusions and outlook}
%
%
%
We are moving towards new frontiers with non-leptonic $B$ decays. Concerning the benchmark decays 
$B^0_d\to J/\psi K_{\rm S}$ and $B^0_s\to J/\psi \phi$, penguins corrections can be
included through control channels, utilising a combined strategy of various channels. In view of the impressive future precision 
for $\phi_d$ and $\phi_s$, this will be crucial for matching the experimental with the theoretical precisions. Future measurements of 
$\phi_s$ may still result in NP effects with more than $5\sigma$ significance. 

In the $B^0_s\to D_s^\mp K^\pm$ system, intriguing CP-violating observables were measured that are complemented through puzzling patterns in the $B^0_s\to D_s^\mp K^\pm$ branching ratios, suggesting NP effects in the decay amplitudes. The pattern of the branching ratios is complemented with analogous phenomena in other decays with similar dynamics. A model-independent framework to include and constrain the corresponding NP effects was developed and can be further applied in the future. 

Puzzling patterns arise also in $B\to\pi K$ decays. An amplitude isospin relations with minimal $SU(3)$ input gives intriguing correlations for the CP asymmetries of the $B^0_d\to \pi^0 K_{\rm S}$ decay, where mixing-induced CP violation plays a key role. These patterns could be accommodated through a modified electroweak penguin sector, which could arise in NP scenarios with extra
$Z'$ bosons.

The current situation raises exciting questions: Could the puzzling patterns in the $B^0_s\to D_s^\mp K^\pm$ and $B\to\pi K$ decays actually be manifestations of the same kind of New Physics? Could there be links with the ``$B$ decay anomalies" in semi-leptonic 
and leptonic rare $B$ decays receiving currently a lot of attention? What about specific NP models and links with collider physics 
and direct NP searches? It will be interesting to shed further light on these topics and questions in the future.

\end{document}